\documentclass[twocolumn,prl,aps,showpacs]{revtex4}
\usepackage{graphicx}
\usepackage{amsmath}

\begin{document}

\title{Hydrodynamics of Superfluid Helium in a Single Nanohole}
\date{\today}
\author{M. Savard, G. Dauphinais and G. Gervais}
\email{gervais@physics.mcgill.ca}
\affiliation{Department of Physics, McGill University, Montreal, Canada, H3A 2T8}

\begin{abstract}
\vspace*{0.2cm}
	The flow of liquid helium through a \textit{single} nanohole with radius smaller than 25 nm was studied for the first time.  Mass flow was induced by applying a pressure difference of up to 1.4 bar across a 50 nm thick $Si_{3}N_{4}$ membrane and was measured directly by means of mass spectrometry.  In liquid He I, we experimentally show that the fluid is not clamped by the short pipe with diameter-to-length ratio $D/L\simeq 1$, despite the small diameter of the nanohole.  This viscous flow is quantitatively understood by making use of a model of flow in short pipes.   In liquid He II, a two-fluid model for mass flow is used to extract the superfluid velocity in the nanohole for different pressure heads at temperatures close to the superfluid transition. These velocities  compare well to existing data for the critical superflow of liquid helium in other confined systems.
	
\end{abstract}

\maketitle

Capillary flow experiments have been conducted for decades in  micron sized channels but only recently has the technology become available to fabricate a cylindrical flow channel of any desired dimension in the nanometer range. Flow experiments across solid-state nanoholes (and nanopores) \cite{DEKKER06} have attracted a lot of attention over the last few years mainly for the detection of macro-molecules \cite{LI03}. The same fabrication technique can readily be used to  study flow of many fluids, whether classical or  quantum in nature.  In this work, we are interested in the transport properties of a quantum fluid, liquid helium, flowing through a single cylindrical nanohole of 22.8 nm radius.  This is motivated by a desire to understand the flow of liquid helium in very small apertures, where ultimately in the one-dimensional limit, the quantum fluid might form  the long sought-after Luttinger liquid \cite{AFFLECK11}.

Liquid helium in constrained geometries behaves differently than in the bulk and has therefore been subject to extensive investigation in porous media such as Vycor \cite{REPPY99}, zeolites \cite{TODA07} and aerogel \cite{CHAN98}, as well as in superfluid films \cite{SAUNDERS98}. The experiment we present here has some advantages over these systems, which are made up of a large number of channels and the signal one extract from them is necessarily averaged over the whole distribution of dimensions and defects present in the material.  The fabrication and detection method we use gives us complete control of the size and shape of the nanohole  (from $\sim 1$ nm to 100's of nm diameter),  whereas porous media typically have fixed dimensions for a given material.
\begin{figure}[hbt]
	\centering
		\includegraphics{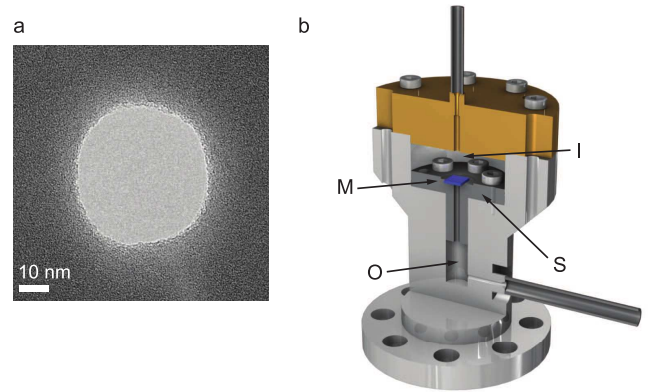}
	\caption{(a) TEM imaging of nanohole.  (b) Schematics of the experimental cell.  The SiN membrane (M) is epoxied to a support (S) to separate the cell between the inlet (I) and outlet (O). }
	\label{fig:cellSchematic}
\end{figure}

Confined  systems have been of great interest for the study of superfluid helium, in particular for the measurement of superfluid critical velocity.  When superfluid helium is forced through a channel, it eventually reaches a critical velocity at which point it begins to dissipate energy, but exactly how a superfluid dissipates energy has been a long standing problem in condensed matter physics.  Critical velocities have been experimentally measured in many systems and were found to vary from mm/s to several m/s, values that are much lower than the Landau critical velocity, $\sim 60 m/s$. This discrepancy and the strong dependence on channel diameter of observed critical velocities were explained by a size-dependent model of quantized vortices in the superfluid.  Many flow experiments found  critical velocities independent of the system size and with a near-linear dependence on temperature \cite{CLOW67}. These were coined  intrinsic critical velocities so as  to distinguish them from the \textquotedblleft extrinsic\textquotedblright size-dependent ones and they have been studied by many groups \cite{ZIMMERMANN98,PACKARD07,AVENEL03} in superfluid flow experiments, for pores in the range $10^{-5}-10^{-6}$m. The currently accepted interpretation of these results is a thermally-activated vortex nucleation process acting as a dissipation mechanism for the superfluid \cite{VAROQUAUX86}.  In this work, the superfluid flow constrained to a  nanometer-sized single channel is studied such that its superfluid velocity could be inferred from the measured mass flow and compared with existing critical veloctity data. To our knowledge this is the first time that a superfluid mass flow was directly detected for an aperture well in the nanometer scale regime.

	Our sample is made from a 50 nm thick low-stress amorphous silicon nitride (Si$_3$N$_4$) layer grown on a 2.7 mm wide square silicon wafer with a  region where the silicon was etched to obtain a rectangular silicon nitride membrane $\sim 30\mu$m wide.  A cylindrical nanohole with $22.8 \pm 0.7$ nm radius was drilled in this membrane  using a field emission TEM (see Fig. \ref{fig:cellSchematic}a).  Confirmation of the radius of the nanohole was accomplished by measuring the mass flow in the gas phase of helium at 20 K in the Knudsen regime. In this regime, a Knudsen effusion model of gas flow was shown previously \cite{SAVARD09} to give an accurate measure (within a few percent) of  the nanohole radius.  This gas flow measurement yielded a radius of $23.1 \pm 0.5$ nm for the sample used in this work.  It was conducted after all data was taken and as such provides strong support that the nanohole size remained unchanged  throughout the experiment.
 
 The sample wafer was epoxy sealed to a support made of Invar alloy separating two reservoirs (inlet and outlet) in an experimental cell designed such that any mass transfer between the two reservoirs is restricted to occur through the nanohole. Capillaries connect the extremities of the experimental cell to a gas handling system such that pressurized helium can be introduced in the cell, flow through the nanohole, and be pumped from the outlet of the cell by a mass spectrometer (see Fig. \ref{fig:cellSchematic}b). We used silver powder packed heat exchangers to condense helium before it enters the inlet of the cell and to ensure a good thermal anchor to the cryostat.  Since the outlet is kept under vacuum, the pressure $P_{in}$ of the liquid helium above the membrane determines the pressure difference $\Delta P = P_{in} - P_{out} = P_{in}$  driving the mass flow through the nanohole.  This mass flow was measured between 1.6 and 3.5 K and up to 1.4 bar pressure difference across the nanohole. A calibrated leak of $4.5\times10^{-3} ng/s \pm 10\%$ was used to calibrate the mass spectrometer before the experiments.

	A typical experiment goes as follows: we first empty both sides of the cell at a temperature well above the helium boiling point so as to ensure that no residual helium is present in either reservoir. The mass spectrometer is then connected to the outlet of the cell to determine a background signal that is treated as an offset to the pressure-driven flow of interest in this study.  This background signal was always found to be below $5\times10^{-4}$ ng/s, which is less than the mass flow measurement presented here by a few orders of magnitude. The next procedure is a cooling of the whole apparatus  below the $\lambda$-transition so that gaseous helium introduced from the gas handling system condenses and fills the heat exchanger and inlet of the experimental cell. Once condensation is achieved, the higher pressure above the membrane forces the liquid helium to flow through the nanohole.  When atoms reach the very low pressures in the lower reservoir, they evaporate and are pumped out to the mass spectrometer.  The mass flow signal is then monitored as the temperature is slowly increased.  This whole procedure is repeated at different pressure difference simply by varying the pressure $P_{in}$ above the nanohole.

\begin{figure}[htb]
	\centering
		\includegraphics{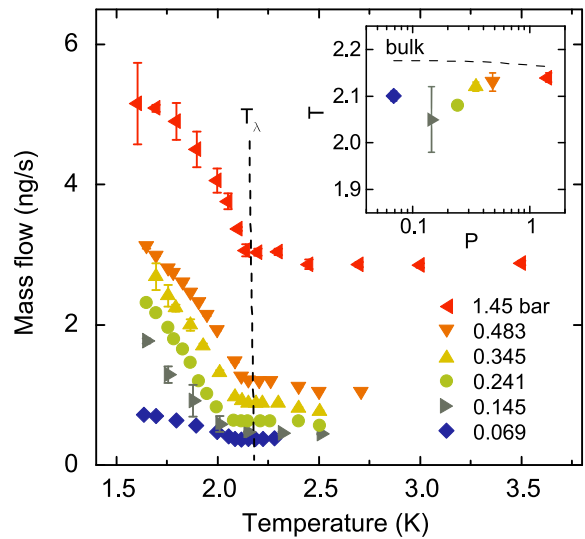}
	\caption{Mass flow of liquid helium in a single nanohole of 22.8 nm radius at pressure differences of 0.069, 0.145, 0.241, 0.345, 0.483 and 1.45 bar respectively from bottom to top. The inset shows the pressure dependence (in bar) of the transition temperature (in Kelvin) for the onset of superfluid mass flow through the nanohole; the bulk superfluid transition  $T_{\lambda}$ is shown by a dashed line.}
	\label{fig:AllFlows}
\end{figure}

The results of several temperature sweeps are presented in Fig. \ref{fig:AllFlows}.  The mass flow is monitored at constant pressure while the temperature of the entire cell is increased  from $\sim 1.6$ K to temperatures above $T_{\lambda}$. At each temperature, the system is  given several minutes to equilibrate thermally until  the mass flow signal reaches a new constant value. The time constant for the system to converge to this new value was found to be  between $\sim$600 and $\sim$1200 s,  depending on temperature.

Focusing on the data at temperatures above $T_{\lambda}$ in Fig. \ref{fig:AllFlows},  our data unambiguously shows a strong dependence of the mass flow on pressure. This means viscous flow of liquid helium  through the nanohole is observed, and by extension that the flow of the normal component should not be ignored in  helium II. Indeed one might naively make the assumption that it should be neglected since channels of the order of $10^{-7}$m are frequently used  as a superleak to clamp the normal component and measure exclusively superfluid signals. Our work shows this is not the case for this nanohole geometry. 
%
%
%

 Modeling of the mass flow must take into account the acceleration of the fluid at the ends of the short pipe, since the diameter-to-length ratio $D/L\simeq 1$ for this nanohole is too large for the infinite-pipe approximation to hold. A derivation for a viscous flow through a  short cylindrical channel was derived by Langhaar \cite{LANGHAAR42} and gives $\Delta P = \frac{32\eta L v}{\rho D^2} + \frac{1}{2}\alpha v^2$, where $\alpha$ is a geometric parameter introduced to take into account the acceleration of the fluid near the ends, and $\eta$,  $\rho$ and $v$ are the dynamic viscosity, density and  average velocity, respectively. Solving for $v$ gives 
\begin{equation}
v = \frac{32\eta L}{\alpha \rho D^2}\left( \sqrt{1+\frac{\alpha \rho D^4}{512 \eta^2 L^2}\Delta P} - 1\right).
		\label{Eq:AvgVelocity}
\end{equation}
From the definition of mass flow  $Q_m = \rho  v A$, with  $A$ the effective area of the nanohole, we obtain, 
\begin{equation}
Q_m = \frac{8\pi\eta L}{\alpha} \left( \sqrt{1 + \frac{\alpha\rho R^4 }{32\eta^2 L^2}\Delta P}- 1 \right).
		\label{Eq:Q_viscous_short_pipe}
\end{equation}
This mass flow equation can be fitted to the data in Fig. \ref{fig:AllFlows}  (at temperatures above $T_{\lambda}$) using the radius $R$ as a varying parameter. 
The strong dependence of Eq. \ref{Eq:Q_viscous_short_pipe} on the radius provides a quantitative test of our short pipe model. In addition, 
the extracted radii can be readily compared to the TEM image taken before the measurements.  The temperature and pressure dependence of $\eta$ and $\rho$  are known from the literature \cite{GOODWIN68}, and  the length $L$ of the short pipe was determined during the sample fabrication process.  The value of the parameter $\alpha =4.7$ was determined in earlier experiments on viscous gas flow in samples with very similar geometry \cite{SAVARD09}. The radii were extracted from all curves in Fig. \ref{fig:AllFlows},  and the average radius is found to be $\bar{R}=20$ nm with a standard deviation $2$ nm. This value is in close agreement with the radius of $22.8 \pm 0.7$ nm extracted from the TEM picture, giving us confidence in the appropriateness of the model used for the normal flow through the nanohole.

 \begin{figure}[hbt]
	\centering
		\includegraphics{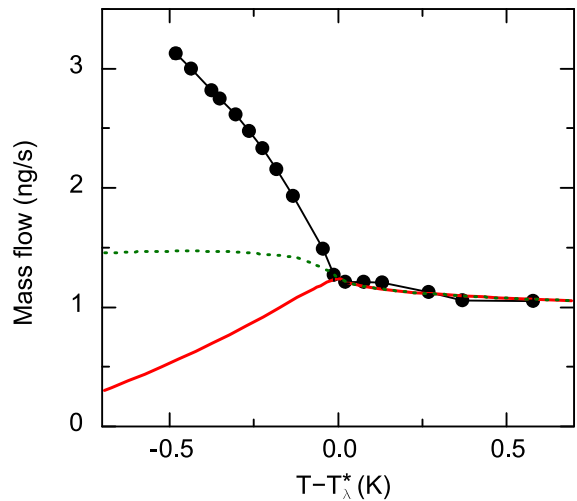}
	\caption{Mass flow measured at a pressure of 483 mbar (solid circle). The green dotted line shows the expected flow if only normal fluid flow was present (no superfluid component). The red curve (second term of Eq. \ref{Eq:Qtotal}) is the viscous flow contribution in the two-fluid model for short pipe using the tabulated normal density. The black line is a guide to the eye.}
	\label{fig:Fit21psi}
\end{figure}

At temperatures below the  $\lambda$-transition, a strong increase in the mass flow is observed at all pressures.  We attribute this increase in mass flow to the onset of the superfluid behavior in the nanohole.  Indeed, as shown on Fig. \ref{fig:Fit21psi} (green dotted line), the mass flow $Q_m$ predicted by Eq. \ref{Eq:Q_viscous_short_pipe} using the tabulated total density of liquid helium and viscosity of the normal component is much smaller than what is  experimentally observed; a superfluid contribution must therefore be included in the modeling of the flow.  As presented in the inset of  Fig. \ref{fig:AllFlows}, this departure of the observed mass flow from the viscous flow prediction occurred at slightly lower temperature ($T^*_{\lambda}$) than the known superfluid transition ($T_{\lambda}$).  While the exact nature of this small shift is unknown, it might be due to thermal or confinement effects in the nanohole. To account for this, we have used in the last two figures the temperature difference  $T-T^*_{\lambda}$ rather than the absolute temperature (T) in order to compare more directly with past experiments. 

 The two-fluid model proposed by Tisza and Landau for flow density of He II  is a natural starting point to model the flow when superfluid and normal helium participate in the mass transport through the nanohole. In this model the total mass current is $\textbf{J}_{total} = \rho_s \textbf{v}_s +\rho_n \textbf{v}_n$,
where $\rho_s$ and $\rho_n$ are densities of superfluid and normal component of He II, respectively, and $\rho=\rho_{s}+\rho_n$. We consider the flow to be only in the axial direction of the nanohole so the total mass flow is given by  $Q^{total}_{m}= J_{total} \pi R^2$.  The normal part of He II can be described with the viscous short-pipe flow from above using the normal helium density in the equation.  The total mass flow is therefore given by
\begin{equation}
Q^{total}_{m} = \pi R^2 \rho_s v_s + \frac{8\pi\eta L}{\alpha} \left( \sqrt{1 + \frac{\alpha\rho_n R^4 }{32\eta^2 L^2}\Delta P}- 1 \right).
		\label{Eq:Qtotal}
\end{equation}

\begin{figure}
	\centering
		\includegraphics{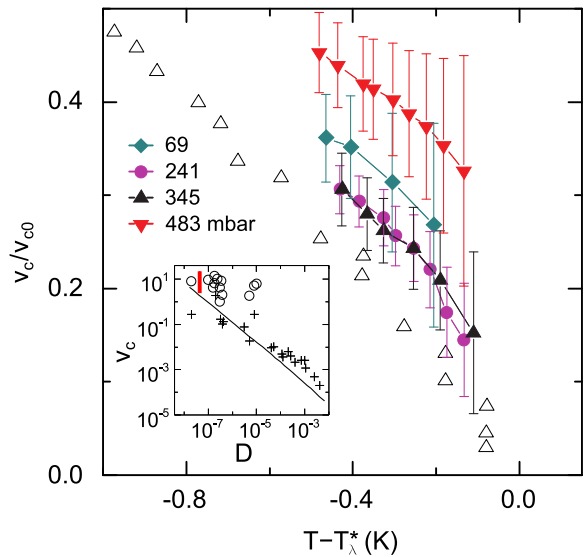}
	\caption{Temperature dependence of the normalized critical velocity. Open triangles (from \cite{ZIMMERMANN98}) are compared to our results (full symbols) at several pressures. (Inset) Critical velocity (in m/s) of superfluid helium from experiments in many different systems \cite{VAROQUAUX06} as a function of channel size (D), in meter.  The red bar shows the range of superfluid velocities inferred from the mass flow through the single nanohole. The open circles are previous results  \cite{VAROQUAUX06} with a temperature-dependent intrinsic critical velocity whereas the crosses are temperature-independent (extrinsic) velocities. The straight line is the Feynman model of critical velocity. 
}
	\label{fig:critVelocity}
\end{figure}

We show in Fig. \ref{fig:Fit21psi} (solid red line) the second term of Eq. \ref{Eq:Qtotal} using the best radius to fit to the He I mass flow. The rapid drop of this term as temperature decreases indicates our measured  signal is decreasingly caused by viscous normal flow. The normal component in the He II region was subtracted from the total flow measured to obtain the superfluid contribution to the mass flow.  From this superfluid mass flow we can infer the average superfluid velocity using the first term on the right-hand side of Eq. \ref{Eq:Qtotal}. We have repeated this procedure for all data sets and  have obtained the  temperature dependence of the superfluid velocity in the nanohole. The critical velocities in other experiments are typically reached at pressure heads of $\sim O(1)$ Pa so given the larger pressure differences applied in our study, we assume the superfluid velocity measured must be critical as well and  can be compared as such to existing data. In the inset of Fig. \ref{fig:critVelocity} we compare the superfluid velocities at 1.7 K to the critical velocities of many other experiments (taken from \cite{VAROQUAUX06}). The open circles are identified as intrinsic critical velocities, whereas the cross symbols are dependent on channel size and follow more closely the Feynman critical velocity model $v_c = \frac{\hbar}{m_4 d}ln(\frac{d}{a_0})$, with $a_0$ the size of the vortex core \cite{VORTEX_CORE}.  Our data (red bar) are consistent with previous results for the intrinsic case.

In previous experiments, it was found that the critical velocity changes almost linearly with a decrease in temperature, and is well characterized by the equation $v_c=v_{c0}(1-T/T_0)$ where $T_0$ and $v_{c0}$ are fitting parameters. In Fig.\ref{fig:critVelocity},  the data from Zimmermann \textit{et al.} \cite{ZIMMERMANN98} are shown and compared to our normalized data for superfluid velocity.  The parameter $v_{c0}$ is the critical velocity extrapolation at $T=0$ K and its value is typically of the order of 1 to 25 m/s \cite{ZIMMERMANN96,ZIMMERMANN98}.  For the nanohole flow, the extrapolation to $T=0$ K yields values of $v_{c0}$ from 8 to 45 m/s. The error on $v_s$ becomes larger in the vicinity of $T_{\lambda}$ because of the uncertainty on the superfluid density.  This latter error prevents us from resolving  the behavior of $v_s$ very near $T_{\lambda}$.  Nonetheless, the behavior of the superfluid velocity as a function of temperature in our nanohole is similar to that observed in larger channels (see Fig. \ref{fig:critVelocity}), albeit the absolute values are the largest ever measured (to our knowledge) in a channel flow experiment.  
%
%
%

Finally, we comment on the possibilities that are being opened up by this work. A theoretical model was constructed that takes into account the flow of liquid
helium through the short pipe. This experiment demonstrates {\it de facto} that mass flow measurements can be performed using liquid helium confined at the nanoscale. Our data are understood in terms of  a two-fluid model modified for the specific geometry of the nanohole short pipe. As the size of the nanohole is reduced, 
and the transport becomes one-dimensional, {\it i.e.} for $D\simeq 1$ nm and $L/D \gtrsim 10$, we expect new flow properties to emerge. In this 1D regime, a breakdown of the two-fluid model is likely to occur,  giving way to the physics of Luttinger liquids which is predicted to form inside the nanopore \cite{AFFLECK11}.


%
%
%
This work has been supported by NSERC (Canada), FQRNT (Qu\'{e}bec),  and CIFAR. We thank J. Hedberg, R. Talbot and J. Smeros for technical assistance,  as well as I. Affleck, W. Mullin and R.B. Hallock  for illuminating discussions. We also thank the McGill Center for the Physics of Materials, the McGill Nanotools microfabrication facility, as well as the hospitality of the Centre for Characterization and Microscopy of Materials (CM)$^{2}$ at \'Ecole Polytechnique de Montr\'eal.

\begin{thebibliography}{25}

\bibitem{DEKKER06}
D. Krapf, M. Y. Wu, R. M.M. Smeets, H. W. Zandbergen, C. Dekker, and S. G. Lemay, Nano Lett. \textbf{6}, (2006).

\bibitem{LI03}
J. Li, M. Gershow, D. Stein, E. Brandin, and J. A. Golovchenko, Nature Mater.  \textbf{2}, 611 (2003).

\bibitem{AFFLECK11}
A. Del Maestro, M. Boninsegni, and I. Affleck, Phys. Rev. Lett. \textbf{106}, 105303 (2011), and references therein.

\bibitem{REPPY99}
G. M. Zassenhaus, and J. D. Reppy, Phys. Rev. Lett. {\bf 83}, 4800 (1999).

\bibitem{TODA07}
R. Toda \textit{et al.}, Phys. Rev. Lett. \textbf{99}, 255301 (2007).

\bibitem{CHAN98}
J. Yoon, D. Sergatskov, J. Ma, N. Mulders, and M. H. W. Chan, Phys. Rev. Lett. \textbf{80}, 1461 (1998).


\bibitem{SAUNDERS98}
J. Ny\'eki, R. Ray, B. Cowan, and J. Saunders, Phys. Rev. Lett. {\bf 81}, 152 (1998), and references therein.


\bibitem{CLOW67}
J. R. Clow and J. D. Reppy, Phys. Rev. Lett. \textbf{19}, 291 (1967).

\bibitem{ZIMMERMANN98}
W. Zimmermann, Jr., C. A. Lindensmith, and J. A. Flaten, J. Low Temp. Phys. \textbf{110}, 497 (1998).

\bibitem{PACKARD07}
Y. Sato, A. Joshi, and R. Packard, Phys. Rev. B \textbf{76}, 052505 (2007).

\bibitem{AVENEL03}
E. Varoquaux, and O. Avenel, Phys. Rev. B \textbf{68}, 054515 (2003).

\bibitem{VAROQUAUX86}
E. Varoquaux, M. W. Meisel, and O. Avenel, Phys. Rev. Lett. \textbf{57}, 2291 (1986) and references therein.

\bibitem{SAVARD09}
M. Savard, C. Tremblay-Darveau and G. Gervais, Phys. Rev. Lett. \textbf{103}, 104502 (2009).

\bibitem{LANGHAAR42}
H. L. Langhaar, J. Appl. Mech. \textbf{9}, A55 (1942).

\bibitem{GOODWIN68}
J. M. Goodwin, Ph.D. Thesis, University of Washington, 1968;
J. S. Brooks, and R. J. Donnelly, J. Phys. Chem. Ref. Data \textbf{6}, 51 (1977).

\bibitem{VAROQUAUX06}
E. Varoquaux, C. R. Phys. \textbf{7}, 1101 (2006).

\bibitem{VORTEX_CORE}
The value of $a_0$ is $\sim 1 - 6$ \r{A} but this uncertainty has negligible effect on Fig. \ref{fig:critVelocity}.

\bibitem{ZIMMERMANN96}
W. Zimmermann, Jr., Cont. Phys. \textbf{37}, 219 (1996).



\end{thebibliography}
\end{document}